\documentstyle[12pt]{article}

\title{Decoherence and the theory of\\ continuous quantum 
measurements\thanks{Published in Uspekhi Fiz. Nauk {\bf 168}, 1017-1035 (1998), \protect\newline
English translation in Physics-Uspekhi 41, 923-940 (1998)}}
     
\author{M.~B.~Mensky \\
{\small P.N.Lebedev Physical Institute of the Russian Academy of Sciences}\\
{\small Moscow 117924 Russia}}
\date{}

\newcommand{\eq}[1]{(\ref{#1})}

\newcommand{\Eq}{Eq.~\eq}

\newcommand{\partderiv}[2]{\frac{\partial #1}{\partial #2}}
\newtheorem{remark}{Remark}

\newcommand{\lo}{\mathrel{\raisebox{- .8 ex}
     {$\stackrel{\textstyle <}{\sim}$}}}

\newcommand{\al}{\alpha}
\newcommand{\tr}{\mbox {Tr}}
\newcommand{\be}{\begin{equation}}
\newcommand{\ee}{\end{equation}}
\newcommand{\ba}{\begin{eqnarray}}
\newcommand{\ea}{\end{eqnarray}}
\newcommand{\ban}{\begin{eqnarray*}}
\newcommand{\ean}{\end{eqnarray*}}

\newcommand{\ra}{\rangle}
\newcommand{\la}{\langle}

\newcommand{\r}{{\bf r}}
\newcommand{\D}{{\Delta}}
\newcommand{\dF}{\delta F}

\newcommand{\Da}{\Delta a}
\newcommand{\ioverh}{\frac{i}{\hbar}}
\newcommand{\k}{\kappa}

\renewcommand{\S}{{\cal S}}
\newcommand{\M}{{\cal M}}
\newcommand{\DE}{{\Delta E}}
\newcommand{\Tlr}{{T_{\rm lr}}}

\begin{document}
\maketitle


Bibliography: 82 references

\vspace{0.5cm}

\begin{abstract}
Decoherence of a quantum system (which then starts to display classical features) results from the interaction of the system with the environment, and is well described in the framework of the theory of continuous quantum measurements (CQM). Reviewed are the various approaches to the CQM theory, and the approach based on the effective complex Hamiltonians is discussed in greater detail. The effective complex Hamiltonian is derived from the restricted path 
integral, which emphasizes the role of information in the dynamics of the system being measured. The complex Hamiltonian is used for analyzing the CQM of energy in a two-level system. Such measurement is demonstrated to be capable of monitoring the quantum transition, and the back effect of monitoring on the probability of transition is analyzed. The realization of this type of measurement by a long series of soft observations of the system is presented.
\end{abstract}

\newpage


\contentsline {section}{\numberline {1}Introduction}{3}
\contentsline {section}{\numberline {2}Measurement of quantum system by its environment}{8}
\contentsline {subsection}{\numberline {2.1}Environment-induced superselection}{9}
\contentsline {subsection}{\numberline {2.2}Models of continuous measurements}{12}
\contentsline {section}{\numberline {3}Phenomenology of continuously measured systems}{14}
\contentsline {subsection}{\numberline {3.1}Approaches to describing continuous measurements}{15}
\contentsline {subsection}{\numberline {3.2}Restricted path integrals (quantum corridors)}{19}
\contentsline {subsubsection}{\numberline {3.2.1}Main principles}{19}
\contentsline {subsubsection}{\numberline {3.2.2}Monitoring of arbitrary observable}{24}
\contentsline {subsubsection}{\numberline {3.2.3}*Uncertainties in continuous measurements}{27}
\contentsline {subsubsection}{\numberline {3.2.4}Features of RPI-based approach}{30}
\contentsline {subsection}{\numberline {3.3}*Derivation of the stochastic equation}{32}
\contentsline {subsection}{\numberline {3.4}*Consistent histories}{34}
\contentsline {section}{\numberline {4}Continuous measurement of discrete energy}{36}
\contentsline {subsection}{\numberline {4.1}Energy measurement in a multilevel system}{38}
\contentsline {subsection}{\numberline {4.2}Monitoring of quantum transition}{42}
\contentsline {subsection}{\numberline {4.3}Realization of continuous measurement of energy}{44}
\contentsline {section}{\numberline {5}Conclusion}{46}

\newpage

\section{Introduction}

Continuous or repeated measurements of quantum systems have been actively discussed over the past decade --- in the first place because this is where the intrinsic features of quantum theory are manifested to the utmost, and secondly because such measurements are steadily gaining practical importance 
\cite{Davies-bk76}-\cite{Zeh-bk96}.
Twenty years ago it was theoretically demonstrated 
\cite{Khalfin58ZenoEng}-\cite{Peres80-Zeno}, 
and later confirmed experimentally \cite{Itano90-Zeno} that repeated measurements of a discrete observable lead to 
freezing of the system in the original state (the so-called quantum Zeno effect). If, however, the accuracy of each of repeated measurements is not high, then their effect on the measured system is not as strong, and the continuously measured system is not frozen \cite{Men-bk83eng,Men-bk93}. Recently it was demonstrated \cite{AudMen97En,AuMenNam97scat} that a soft continuous quantum measurement (CQM) is capable of monitoring a quantum transition, and an experiment was proposed in which the transition between 
two atomic levels induced by resonant radiation is monitored through observing a series of electron scatterings by the atom. Actually, instead of electron scattering one could use a series of short and weak interactions of the atom with any auxiliary system. Other quantum phenomena can be monitored in a similar way. Thus, soft continuous quantum measurements offer a new tool for experimental study of quantum processes, capable of giving an insight deeper than previously deemed possible.

We shall outline these new possibilities at the end of this review, and will start it with the theoretical analysis of quantum measurement in general, and CQM in particular. We are going to give a thorough analysis of the phenomenon of decoherence of quantum system --- the process through which the system acquires classical features, and the environment stores information about the system [12]. The phenomenon of decoherence is associated with any measurement, 
and plays the decisive role in the dynamics of a system subjected to repeated or continuous quantum measurements.

Using a simple example we shall see how the measurement of a quantum system occurs in the process of interaction between the system and its environment, and why this interaction inevitably leads to decoherence of the measured system. We shall briefly discuss two models of quantum diffusion --- that is, the continuous monitoring of the coordinate of the quantum particle. Then we are going to present different phenomenological approaches to CQM, which 
enable studying them without resorting to particular models of the surrounding (measuring) medium.

In special detail we shall discuss the phenomenological approach based on the restricted integrals over paths (quantum corridors), which effectively reduces to the Schr\"odinger equation with complex Hamiltonian. Subsequently we shall use this approach for analyzing the continuous measurement of the energy of a multilevel system and --- in greater detail --- a two-level one. We shall prove that such measurement is capable of monitoring the quantum transition (Rabi oscillations).

\centerline{$\bullet$}

The relationship between quantum and classical descriptions of physical processes has been actively discussed since the early days of quantum mechanics, and still is. This issue is highly complicated because of the fundamental differences between the quantum and classical representations of physical systems and phenomena. The main distinction is apparently the quantum theoretical principle of superposition for the states of ``corpuscular" 
systems, for which in classical physics superposition is not possible. Because of this, it is not quite true that classical theory gives an approximate description of something that is more precisely rendered in terms of quantum theory. Niels Bohr believed that a complete description of physical phenomena ought to include classical elements in addition to quantum representation.

Quantum mechanics assumes that, along with any two states of a quantum system $|\psi_1\ra$, $|\psi_2\ra$, there exists also their quantum mechanical (also known as coherent) superposition $|\psi\ra=c_1|a_1\ra + c_2|a_2\ra$. If, however, the states $|\psi_1\ra$, $|\psi_2\ra$ differ considerably --- are said to be classically distinctive --- for example, relate to the states of an elementary particle localized at points far from each other, then their superposition is not observed under ordinary circumstances. Such states may be referred to as non-coherent.

To be more precise, a superposition of macroscopically distinctive states can be realized; but to have it survive for some time the system must be completely isolated. Even a very slight interaction with the environment will very quickly reduce the superposition $|\psi\ra$ to one of the stable states $|\psi_1\ra$ or $|\psi_2\ra$, with only the respective probabilities $|c_1|^2$ and $|c_2|^2$ known in advance. Such conversion is the simplest example of decoherence. The fathers of quantum mechanics 
referred to this process as reduction or collapse. Today its nature is well understood, and studied is the dynamics of this process, its development in time.

A couple of decades ago the process of decoherence was mostly of academic interest. Currently, however, the experimental techniques in the field of, for example, quantum optics \cite{WallsMilb-bk94quOpt}, have become so much refined that it has become possible to observe the formation of superposition of macroscopically distinctive states and the subsequent process of decoherence \cite{Knight92cat,Knight94cat}. Moreover, the processes of decoherence must necessarily be included into the correct description of 
quantum systems interacting with the environment (open systems).

In particular, the processes of decoherence are important for the theory and practice of quantum computers \cite{Unruh95quCmp,Landauer96quCmp}. Quantum computer is a device capable of performing parallel computations by operating with a quantum superposition involving an enormous number of terms.

In quantum computers (whose elements have already been realized in practice) decoherence plays a dual role. In the course of calculations, decoherence is a harmful process, since the superposition has to be prevented from falling apart. However, when the computation is complete, its result must be retrieved from the computer and represented in the classical (that is, stable and steady) form. This is accomplished by the appropriate measurement of the state of computer as a 
quantum system. In other words, the computer is made to interact with a special device which acts as a measuring system. This device measures certain parameters of the state of the computer, thus causing decoherence of this state. In this case the decoherence is brought about deliberately.

Hypothetical quantum computers are devices in which decoherence plays a crucial role. It is of no less importance, however, for many other quantum devices. The phenomenon of decoherence occurs whenever the system interacts or is made to interact with its environment, and the state of the system has some impact on the state of the environment. By observing the state of the environment one can then gain some information regarding the state of the system. Accordingly, 
interaction of the system with its environment may be interpreted as the measurement of the system. The information about the system is recorded in the environment. Then we say that a measurement of the quantum system has taken place. The environment that performs this measurement can be created on purpose (measuring device or measuring medium), although in many cases it exists beyond the experimenter's discretion, and often plays an undesirable role, leading to a special kind of dissipation.

It is important that such interaction with the environment inevitably modifies the state of the quantum system, causing decoherence. It turns out, however, that the behavior of the quantum system being measured can be described with due account for its decoherence, without the need for explicit description of the measuring medium. The back effects of the environment are taken into account implicitly. This means that the system in question is treated as an open system, and its 
evolution is described phenomenologically.

The phenomenological theory of continuously measured quantum systems is an extension of the conventional quantum mechanics and much augments its capabilities. This theory is closed and intrinsically elegant.

In this way, the answers to the questions ``How does a quantum measurement occur?", and ``How does a continuously measured system behave?" brings us to the theory of open continuously measured quantum systems.

\centerline{$\bullet$}

From the general course of quantum mechanics we know that the measurement on a quantum system obeys von Neumann's reduction postulate, which in the simplest case is represented by the scheme 
$$
c_1|1\ra + c_2|2\ra \rightarrow
\left\{
\begin{array}{ll}
|1\ra, \quad p_1=|c_1|^2 \\
|2\ra, \quad p_2=|c_2|^2
\end{array}
\right.
$$
In this scheme, $|1\ra$ and $|2\ra$ are states corresponding each to a certain outcome of the measurement. According to the reduction postulate, one or the other measurement output are random quantities with respective probabilities $p_1$ and $p_2$, and the system assumes the corresponding state. This instantaneous transition, which cannot be described by the Schr\"odinger equation, is known as reduction or collapse of the state of the system. Mathematically, 
reduction can be described as the projection of the initial state vector on to the subspace of vectors proportional to either $|1\ra$ or $|2\ra$.

In the more general case, the measurement is described by a set of projectors $P_i$, where subscript $i$ numbers the alternative outcomes of the measurement. If the measurement results in the $i$th alternative, the initial state $|\psi\ra$ after the measurement will go into state $|\psi_i\ra=P_i|\psi\ra$ --- that is, the reduction of state is described by the corresponding projector. The probability that the measurement will result in the $i$th alternative is\footnote{Usually the state after the measurement is described in terms of the normalized vector. We, however, prefer using vector $|\psi_i\ra$, since the reduction then is represented in a very straightforward manner, and the norm of the resulting vector is equal to the probability of the respective outcome of the measurement. This approach is especially convenient for describing repeated measurements.}
\be
p_i=||\psi_i||^2=\la\psi_i|\psi_i\ra=\la\psi_i|P_i|\psi\ra.
\label{NeumannProj}\ee

This most simple scheme of description of measurement raises a number of questions, the first of which is how and why the collapse takes place. The answer to this question has been sought by many authors (see for example 
\cite{Mandelstam50eng,Blokhintsev-bk87eng}, 
\cite{Kadomtsev94eng}-\cite{Kadomtsev-bk97eng} 
and an excellent review of the topical literature in the book \cite{Zeh-bk96}). Very clear analysis of the mechanism leading co collapse has been given in papers by Zurek \cite{Zurek81,Zurek82}, where it has been termed environment-induced superselection. We shall briefly touch upon this issue in Section~\ref{Sect-superselect}.

Another question arises when we consider a series of successive quantum measurements each of which is described by von Neumann's reduction. How does a system behave when subjected to a series of such instantaneous measurements? How does this behavior change if the intervals between the successive measurements tend to zero, so that the measurement becomes continuous? This paper for the most part is devoted to the discussion of these questions.

The analysis of these problems reveals that repeated von Neumann measurements of the observable with discrete spectrum (for example, measurement of the energy of a multilevel system) lead to suppression of quantum transitions. As a result of continuous measurement, the system is completely frozen at one point of the spectrum. This phenomenon has attracted much interest and became known as Zeno quantum paradox (or effect) 
\cite{Khalfin58ZenoEng}-\cite{Peres80-Zeno}.

Zeno paradox indicates that a continuous measurement may lead to trivialization, to the disappearance of dynamics. But is it always the case? The answer is negative. In the first place, the system is not frozen if the measured observable has a continuous spectrum. Secondly, the dynamics remains nontrivial even in case of discrete spectrum as long as the measurement is soft (not too accurate). From this standpoint, a too accurate measurement 
of quantum system is not advantageous, and this is a manifestation of the paradoxical nature of quantum mechanics.

The dynamics of a quantum system subjected to continuous measurement is a new type of dynamics which is more general than that described by the conventional Schr\"odinger equation. It includes dissipation due to the effects of the environment. It is the dynamics of open quantum systems.

\centerline{$\bullet$}

There are different approaches to the description of open (continuously measured) quantum systems. Later on we are going to discuss these in greater detail; at this point we shall just enumerate some of them:
\begin{itemize}
\item The model of measurement which includes the principal quantum system $\cal{S}$, its environment (or measuring device) $\cal{M}$, and the interaction between them.
\item The equation for the density matrix (master equation) of the system $\cal{S}$, obtained after summation over the degrees of freedom of the environment $\cal{M}$ (a special case of the Lindblad equation).
\item Restricted path integrals (quantum corridors), which can be reduced to the Schr\"odinger equation with complex Hamiltonian.
\item The stochastic Schr\"odinger equation.
\end{itemize}

Finally, let us make two conceptual remarks.

The theory of open continuously measured quantum systems throws new light on the old question whether or not quantum mechanics is a closed theory. The answer is affirmative if we are considering the Feynman formulation of quantum mechanics which is extremely rich in ideas. This ideological diversity allows us not to introduce the theory of quantum measurements as a special independent postulate, but rather derive it from the Feynman formulation of quantum mechanics.

The phenomenology of continuously measured quantum systems leads to the conclusion of the dynamic role of information in the following sense. Of course, the dynamics of the measured system is determined by the nature of the measuring medium and its interaction with the system. All essential features of the dynamics, however, can be reconstructed using only the information about the system that is recorded in the environment. It is information that determines the dynamics.

\section{Measurement of quantum system by its environment}
\label{Sect-meas}

In this section we shall consider the physically more obvious ``straightforward" descriptions of the measurement of quantum system, leaving for the next section the more abstract phenomenological approaches which have the advantage of being universal and independent of the model used.

\subsection{Environment-induced superselection}
\label{Sect-superselect}

As already mentioned, the behavior of quantum system subjected to measurement (idealized, of course) is described by von Neumann's reduction postulate \cite{vonNeumann55eng}. Let us consider very schematically the physical nature of the reduction (collapse) of state. To illustrate the main idea it will suffice to analyze the simplest measurement amounting to the choice between two alternatives. Assume, for example, that measured is the observable $A$ which may take on one of the two values $a'$, $a''$. Then, as a result of the measurement, the system, with the appropriate probability, will go over into one of the eigenstates of the observable:
\be
|\psi\ra=c'|a'\ra + c''|a''\ra \rightarrow
\left\{
\begin{array}{ll}
|a'\ra, \quad p_1=|c'|^2 \\
|a''\ra, \quad p_2=|c''|^2
\end{array}
\right.
\label{vonNeumann2}\ee

The same change in terms of the density matrix is expressed as the transition of the density matrix $\rho_0 = |\psi\ra\la\psi|$ of pure state into the density matrix of mixed state
\be
\rho_0 =\left(
\begin{array}{cc}
|c'|^2 & c' c''^{*}\\
c'^{*} c'' & |c''|^2
\end{array}
\right)
\rightarrow
 \left(
\begin{array}{cc}
|c'|^2 & 0 \\
0       & |c''|^2
\end{array}
\right) = \rho.
\label{diagonalization}\ee
The resulting density matrix $\rho$ contains the same information as the right-hand side of \eq{vonNeumann2}, including information about the transition probabilities.

The physical process leading to the transition \eq{diagonalization} is called decoherence. It results in the conversion of a superposition of a set of states to the mixture of the same states. The mark of decoherence is the disappearance of the nondiagonal elements of the density matrix:
\be
\la a' |\rho |a''\ra = \la a'' |\rho |a''\ra = 0.
\label{off-diagonal}\ee

After the measurement, the superposition of states $|a'\ra$ and $|a''\ra$ is no longer possible, and only one of them may exist[[ without admixture of the other]]. We speak then of superselection - prohibition of the superposition of states from a given set of subspaces (in our example it is two one-dimensional subspaces).

Now let us consider the physical mechanism which leads to decoherence and superselection. This mechanism consists in the interaction of the measured system with its environment (measuring device) according to the scheme:

\quad

\framebox{System} $\leftrightarrow$ \framebox{Environment}

\quad

\noindent
Interaction leads to entanglement, to quantum correlation of the two systems, so that the state of one of them contains information about the state of the other. This is how this occurs.

As already mentioned, a situation equivalent to the measurement of a quantum system is often encountered even in case when the measurement was not aimed  by the experimenter. Nevertheless, for the sake of simplicity we shall speak of the device $\M$ that measures the system $\S$. Assume that the device prior to the interaction occurs in the state $|\phi_0\ra$. Interaction between two systems is referred to as measurement when it results in a specific 
correlation between these two systems, so that the information about one system is recorded in the state of the other. In our case, the interaction between the two subsystems must translate the state $|a'\ra\; |\phi_0\ra$ of the compound system into $|a'\ra\; |\phi'\ra$, and the state $|a''\ra\; |\phi_0\ra$ into $|a''\ra\; |\phi''\ra$. Then the state of the device after the measurement will tell us about the state of the measured system.

Assume that prior to the measurement the composite system, comprising $\S$ and $\M$ as subsystems, occurs in the state $|\Psi_0\ra=|\psi\ra\; 
|\phi_0\ra$, where $|\psi\ra$ is the state of the measured system discussed above. Then the interaction between the two subsystems will result in the following change of state of the composite system:
\be
|\Psi_0\ra=|\psi\ra\; |\phi_0\ra
               = (c'|a'\ra + c''|a''\ra)\; |\phi_0\ra 
               \rightarrow
c'|a'\ra\; |\phi'\ra + c''|a''\ra\; |\phi''\ra=|\Psi\ra.
\label{intertwining}\ee

Observe that, by contrast to the reduction of states \eq{vonNeumann2} or \eq{diagonalization}, the transition \eq{intertwining} occurs due to the conventional quantum mechanical evolution, and can be described by the Schr\"odinger equation. This evolution results in the state $|\Psi\ra$ of the composite system, in which the subsystems $\S$ and $\M$ are entangled, in other words, quantum correlation is established between them.

Constructing the density matrix $|\Psi\ra \la\Psi |$ of the composite system after the measurement and calculating its trace with respect to all degrees of freedom of the measuring device, we obtain the density matrix of the measured system:
\ba
\lefteqn{\rho = \tr_{\phi} |\Psi\ra \la\Psi |} \\
&&= |c'|^2\, |a'\ra\la a' | + |c''|^2\, |a''\ra\la a'' | 
+ \la\phi'' |\phi'\ra \; c' c''^{*}  |a'\ra\la a'' |
+ \la\phi' |\phi''\ra \; c'^{*} c''  |a''\ra\la a'|. \nonumber
\label{intertwin-matrix}\ea
In this expression the off-diagonal matrix elements are nonzero. The following analysis indicates, however, that in reality they are negligibly small, and condition \eq{off-diagonal} is satisfied to a high degree of accuracy.

In this expression the off-diagonal matrix elements are nonzero. The following analysis indicates, however, that in reality they are negligibly small, and condition \eq{off-diagonal} is satisfied to a high degree of accuracy.

A large (macroscopic) number of degrees of freedom is prerequisite for any measuring device, as is the fact that its states corresponding to different outcomes of the measurement (in our case $|\phi'\ra$ and $|\phi''\ra$) are ``macroscopically distinctive". This means that the corresponding wave functions depend on very many variables, and exhibit different functional dependences on the large number of these variables. The scalar product of such wave functions is practically equal to zero 
(to be more precise, it is exponentially small with the exponent of the order of -10$^{23}$). The reason is that the scalar product is an integral with respect to an enormous (macroscopic) number of variables. Even if the integral with respect to each separate variable is a little less than one, the total multiple integral will be close to zero. Hence, to a high degree of accuracy we have
\be
\la\phi' |\phi''\ra = \la\phi'' |\phi'\ra = 0.
\label{macro-dif}\ee
As a result, the off-diagonal terms of the density matrix vanish, and it becomes
\be
\rho = \tr_{\phi} |\Psi\ra \la\Psi |
= |c'|^2\, |a'\ra\la a' | + |c''|^2\, |a''\ra\la a'' |
\label{induced-diag}\ee
in accordance with \Eq{off-diagonal}. In this way, the measurement leads to decoherence or superselection. The nature of this phenomenon has rather long been understood (see, for example, the excellent papers 
\cite{Zeh70}-\cite{JoosZeh85deco}). 
Zurek furthered the analysis of this phenomenon and aptly christened it ``environment-induced superselection" \cite{Zurek82}.

We did not go into the details of the interaction between the measured system and its environment leading to transition \eq{intertwining}. The analysis of relevant models (see book \cite{Zeh-bk96} and references therein) reveals that decoherence arises (that is, the off-diagonal terms vanish) exponentially fast in accordance with the expression
\be 
\left|\la a''|\rho(t)|a'\ra\right|^2 \sim e^{-\kappa(a''-a')^2 t}. 
\label{exp-time-deco}\ee 
This occurs as more and more degrees of freedom of the environment get entangled with the measured system. As follows from \Eq{exp-time-deco}, the characteristic time of decoherence $t_d$ is inversely proportional to the squared difference between the measured values of the observable,
\be 
t_d\sim (a''-a')^{-2}. 
\label{deco-time}\ee
By the way, this explains why it is practically impossible to realize a superposition of two states of a particle which are localized at points far from each other. Even if such superposition were to arise, it would very quickly suffer decoherence owing to the interaction with the environment from which it cannot be completely isolated.

\subsection{Models of continuous measurements}
\label{QuDiffusion}

The physically most intelligible approach to the description of quantum measurements, including continuous measurements, relies on some model of the process of measurement in accordance with the general scheme outlined in Section 2.1. The model must include the principal quantum system $\cal{S}$, its environment $\cal{M}$, and the interaction between them. If the macroscopic number of degrees of freedom of the environment is explicitly taken into account, then the model becomes rather complicated.

Proposed was a large variety of models of quantum measurements (see, for example, in the chronological order, 
\cite{Brag67eng,BragBk70eng,Zeh70,Zeh73,BragVor74eng,Zurek82,CaldeiraLegg83,
WallsMilburn85,WallsColMilburn85,CavesMilb87,Peres87cont,Vorontsov-bk89eng,
BragKhal-bk92,KonMenNamiot93,Zeh-bk96}). 
Without going into the details, we are going to describe and compare two models of one and the same continuous measurement which consists in monitoring the coordinate of a pointlike particle. The first of these models was proposed as part of a theory of quantum diffusion \cite{CaldeiraLegg83}. Consider a pointlike quantum particle interacting with the 
atoms in crystal lattice as shown in the following diagram of decoherence by crystal modes:
\quad

\noindent {\samepage
{\large\bf Decoherence by crystal modes}\\ 
Caldeira \& Leggett 1983\nopagebreak

\begin{picture}(300, 120)(0,0)
\thicklines
\multiput (0,0)(20,0){15}{\circle*{3}}
\multiput (0,20)(20,0){15}{\circle*{3}}
\multiput (0,40)(20,0){15}{\circle*{3}}
\multiput (0,60)(20,0){15}{\circle*{3}}
\multiput (0,80)(20,0){15}{\circle*{3}}
\multiput (0,100)(20,0){15}{\circle*{3}}
\put (0,27){\line(3,1){23}}
\put (23,34.66){\line(1,1){17}}
\put (40,51.66){\line(1,1){21}}
\put (61,72.66){\line(1,-1){19}}
\put (80,53.66){\line(5,1){41}}
\put (121,61.87){\line(2,-1){41}}
\put (162,41.37){\line(1,2){17}}
\put (179,75.37){\line(2,-1){21}}
\put (200,64.87){\line(2,1){19}}
\put (219,74.37){\line(3,-1){18}}
\put (237,68.37){\line(1,1){21}}
\put (258,89.37){\line(2,-3){7}}
\put (265,78.87){\line(3,1){15}}
\end{picture}
}

\quad

The ``particle" in our context is the measured system $\S$, and the ``crystal" is its environment $\M$. Such interaction results in the correlation between the coordinate of particle and the state of crystal (in other words, the state of phonons that represent the motion of crystal atoms). The particle then suffers decoherence, its state is described by the density matrix, and the time evolution is described by equation\footnote{A more general equation can be found in Ref.~\cite{CaldeiraLegg83}. We give a simplified expression whose validity is subject to certain restrictions on the parameters of the system.}
\be
\dot\rho = -\frac{i}{\hbar} [H,\rho]
          - \frac 12 \kappa [{\bf r}\, [{\bf r}\, \rho]],
\label{qu-diffusion}\ee
where the coefficient
\be
\kappa=\frac{2\eta kT}{\hbar^2} 
\label{kappaCalLeg}\ee
depends on the temperature of the crystal $T$ and the damping coefficient $\eta$, the same as enters the classic equation of motion of particle in a medium
\be
m\ddot \r + \eta\dot \r + V'(\r) = F(t)
\label{class-diffus}\ee

Another model of motion of particle through a medium was specially designed for describing the continuous measurement of the particle's coordinate \cite{KonMenNamiot93,Men97timeResol}. This model assumes that the interaction with the particle excites the internal degrees of freedom of the atom, and the decoherence occurs through interaction with these degrees of freedom rather than with the modes corresponding to the displacement of atoms. This model can be schematically represented by the 
following diagram of decoherence by internal structure of atoms:

\quad

\noindent {\samepage
{\large\bf Decoherence by internal structure of atoms}\\ 
Konetchnyi, Mensky \& Namiot 1993\nopagebreak

\begin{picture}(300, 120)(0,0)
\thicklines
\multiput (0,0)(20,0){15}{\circle{6}}
\multiput (0,20)(20,0){15}{\circle{6}}
\multiput (0,40)(20,0){15}{\circle{6}}
\multiput (0,60)(20,0){15}{\circle{6}}
\multiput (0,80)(20,0){15}{\circle{6}}
\multiput (0,100)(20,0){15}{\circle{6}}
\put (0,27){\line(3,1){23}}
\put (23,34.66){\line(1,1){17}}
\put (40,51.66){\line(1,1){21}}
\put (61,72.66){\line(1,-1){19}}
\put (80,53.66){\line(5,1){41}}
\put (121,61.87){\line(2,-1){41}}
\put (162,41.37){\line(1,2){17}}
\put (179,75.37){\line(2,-1){21}}
\put (200,64.87){\line(2,1){19}}
\put (219,74.37){\line(3,-1){18}}
\put (237,68.37){\line(1,1){21}}
\put (258,89.37){\line(2,-3){7}}
\put (265,78.87){\line(3,1){15}}
\end{picture}
}

\quad

This model also leads to equation of the form of \Eq{qu-diffusion}, but the coefficient
\be
\kappa = \frac{2}{\lambda^2\tau}
\label{kappa-AuMeNa}\ee
in this case will depend on the distance $\lambda$ at which the atom reacts to the particle, and the relaxation time $\tau$ of the atom excited by the passage of the particle. Under certain conditions, this second mechanism of decoherence will prevail.

For our future discussion it is important that both models lead to the equation \eq{qu-diffusion} for the density matrix, in which the effect of decoherence is represented by the double commutator of the density matrix with the measured observable (which in our case is the position of the particle). As will be demonstrated in Section~\ref{Sect-phenomen}, such equation, which represents the phenomenological description of continuous quantum measurement, can be derived without using the model of interaction.

\section{Phenomenology of continuously measured systems}
\label{Sect-phenomen}

As demonstrated in Section~\ref{Sect-meas}, the behavior of a quantum system subjected to continuous measurement can be derived by considering the model of the environment. It is possible, however, to do without such model, taking the effects of the environment into account implicitly in accordance with the scheme

\quad

\framebox{System} $\leftrightarrow$ \\

We shall discuss different approaches to this phenomenological description of continuous quantum measurements (CQM), paying special attention to the method of restricted path integrals.

\subsection{Approaches to describing continuous measurements}

A continuous quantum measurement can be approximated by a series of repeated instantaneous measurements, each of which is described by the von Neumann projector. Evolution of the system between two measurements is described by the Schr\"odinger equation, or (which is equivalent) by the unitary evolution operator. This gives us the following law of evolution between the times $t_0$ and $t=t_N$:
\ba
|\psi(t)\ra = U(t_N,t_{N-1})P_{i_{N-1}}U(t_{N-1},t_{N-2})\dots \nonumber\\
\dots U(t_3,t_2)P_{i_2} U(t_2,t_1)P_{i_1} U(t_1,t_0) |\psi(t_0)\ra
\label{serialMeas}\ea
The result of a series of measurements is represented by a sequence of numbers $\{ i_1, i_2, \dots , i_{N-1} \}$. We see that the evolution of the measured system depends on these numbers --- that is, on the result of the series of measurements. The description can be made more realistic if we represent the instantaneous measurement not by the von Neumann projectors, but rather by their generalizations --- positive operators. The spectrum of possible measurement outputs can then be made continuous.

Passing to the limit from this formalism for repeated measurements, one can obtain a tool for studying continuous quantum measurements (see, for example, Refs~\cite{BarchLanzProsp82,Men-bk83eng,Khal88-ZenoEng}). The resulting approach is phenomenological and does not require an explicit model of the measuring medium. The continuously measured system is then assumed to be open, and the effects of the environment are taken into account in an implicit way.

There are quite a few phenomenological approaches that lead straight to the continuous measurement, without first representing it as a repeated measurement with subsequent passage to the limit. Let us list the main of these.
\begin{itemize}
\item Equation for the density matrix (master equation) of a continuously measured system $\cal{S}$ can be represented in the form
\be
\dot\rho = -\frac{i}{\hbar} [H,\rho]
          - \frac 12 \kappa [A, [A, \rho]], 
\label{LindbladEq}\ee
where $H$ is the Hamiltonian of the measured system, $A$ is the continuously measured observable of the system --- that is, the observable the information whereof is ``recorded" in the state of the environment, and constant $\kappa$ characterizes the strength of interaction between the measured system and its environment. Equation \eq{LindbladEq} is a special case of the Lindblad equation \cite{Lindblad76} which describes a more general class of open systems. In Section~\ref{QuDiffusion} we saw that the equation of this 
form can be derived from the model of interaction between the system and its environment. In Ref.~\cite{Lindblad76}, however, the Lindblad equation has been derived without using any particular model, under assumption of the Markovian nature of the process. Interpretation of the special case \eq{LindbladEq} of the Lindblad equation as describing a continuous measurement can only be corroborated either by the models of measurement, or with the aid of restricted path integrals (see next paragraph).

\item The restricted path integrals or quantum corridors \cite{Men-bk83eng,Men-bk93} reduce the description of continuously measured system $\cal{S}$ to the Schr\"odinger equation with complex Hamiltonian,
\be 
|\dot\psi\ra = \left[ -\frac{i}{\hbar} H 
- \kappa\big( A-a(t)\big)^2 \right] |\psi\ra. 
\label{SchroedComplHam}\ee 
By $a(t)$ here we denote the value of observable $A$ at the time $t$ found in the course of continuous measurement. Thus, unlike the Lindblad equation, this description of continuous measurement is selective: it takes into account the result of the measurement --- that is, the information recorded in the environment (despite the fact that this approach does not rely on any model of the environment). This is the information approach, which assumes that the dynamics of the measured 
system is determined not by the particulars of the environment, but rather by the information recorded in the environment. Equation~\eq{LindbladEq} can be derived from \Eq{SchroedComplHam} by carrying out summation over all possible curves $a(t)$ in a certain way. This approach will be discussed in detail in Section~\ref{RPI}.

\item The stochastic Schr\"odinger equation for a continuously measured system
\be 
d|\psi\ra = \left[ -\frac{i}{\hbar} H 
- \kappa\big( A-\la A\ra\big)^2 \right] |\psi\ra dt 
+ \sqrt{2\kappa} \big( A-\la A\ra\big)  |\psi\ra dw 
\label{StochEq}\ee 
can be derived \cite{PreOnoTam96} from the equation with complex Hamiltonian \eq{SchroedComplHam}. In the stochastic equation, $w$ is a random function of the type of Brownian walk (white noise), which characterizes the effects of the environment. The differential of this variable, which enters this equation, satisfies the condition $dw^2=dt$, which reflects the properties of Brownian motion (the mean deviation is proportional to square root from the time interval). Proposed have also been other stochastic equations \cite{Gisin89stochast}-\cite{DiosiGisinHallPers95stochast} 
leading to the Lindblad equation~\eq{LindbladEq}. The problem is that the stochastic equation cannot be unambiguously derived from the Lindblad equation. The advantage of \Eq{StochEq} is that it follows from the equation with complex Hamiltonian~\eq{SchroedComplHam}, which in turn can be derived from first principles. In Ref.~\cite{GhirardiRimWeb86} and subsequent publications of the same authors, the noise in the stochastic equation was interpreted in a different 
fashion --- not as an effect of the environment, but rather as an independent fundamental physical process called spontaneous localization.
\end{itemize}

\newpage

The relationship between different phenomenological approaches to continuous quantum measurements can be illustrated by the following table:

\quad

\noindent{\large\bf Quantum mechanics of open measured systems}\\[5mm]
\begin{tabbing}

\parbox{7cm}{\underline{Selective description}}
\hspace{0.4cm}\= \hspace{0.8cm}\=
\parbox{4.5cm}{\underline{Nonselective description}}
\\ \samepage
\hspace{2cm}  $\Downarrow$
\> \>
\hspace{2cm}  $\Downarrow$
\\ 
\framebox{%
\parbox{7cm}{{\bf Feynman Quantum Mechanics}:\\Feynman 1948}
}
\> $\searrow$ \>
\\ 
\hspace{2cm}  $\downarrow$
\> \>
\hspace{2cm}
\\ 
\framebox{%
\parbox{7cm}{{\bf Restricted Path Integrals}:\\
             Mensky 1979}
}
\> $\rightarrow$ \>
\framebox{%
\parbox{4.5cm}{{\bf Influence functional}:\\ Feynman \& Vernon 1963}
}
\\ 
\hspace{2cm}  $\downarrow$
\> \>
\hspace{2cm}
\\ 
\framebox{%
\parbox{7cm}{{\bf Complex Hamiltonians}:\\
Golubtsova \& Mensky 1989\\
Mensky, Onofrio \& Presilla 1991}
}
\> $\searrow$ \>
\hspace{2cm}  $\downarrow$
\\ 
\hspace{2cm}  $\downarrow$
\> \>
\hspace{2cm}
\\ 
\framebox{%
\parbox{7cm}{{\bf Stochastic Equations}\\
Diosi 1989\\ Gisin 1989\\ Belavkin 1989}
}
\> $\rightarrow$ \>
\framebox{%
\parbox{4.5cm}{{\bf Master equation}:\\ Lindblad 1976}
}
\end{tabbing}

In this table we emphasize the distinction between the selective and nonselective descriptions of continuous quantum measurements. The selective description is more detailed. It portrays the evolution of the measured system for only one out of many alternative results of measurement. In this description the state of the measured system remains pure (if, of course, it had been pure prior to the measurement). In the equation with complex Hamiltonian, the alternative is given by the 
function $a(t)$, which has a straightforward physical meaning --- it is the result of monitoring the observable $A$. In the stochastic equation, the alternative is given by the random function $w$.

Nonselective description represents evolution of the measured system irrespective of the measurement readout. This description takes into account all possible readouts, and the actual readout is assumed to be not known. Accordingly, the nonselective description (equation for the density matrix, or master equation) can be derived from the selective description by carrying out summation with respect to the alternatives (see, for example, Ref.~\cite{Men94MasterEq}). The transition 
back from nonselective (less detailed) to selective (more detailed) description is ambiguous and requires additional assumptions for removing the uncertainty. The logical links between the approaches, the possibilities for deriving one approach from another, are shown on the diagram by arrows (see also Section~\ref{Sect-deriv-stochast}).

\subsection{Restricted path integrals (quantum corridors)}
\label{RPI}

Let us consider in greater detail the method of restricted path integrals (RPI) for describing continuous quantum measurements. In this approach, the open continuously measured system is described (as opposed to the closed system) not by a single unitary evolution operator, but rather by a whole family of partial evolution operators (propagators), in accordance with the numerous alternative measurement readouts. Each result of the measurement defines one 
channel of quantum evolution with the aid of the relevant partial propagator. And it is all these channels taken together that give a complete description of the dynamics of the open continuously measured system.

\subsubsection{Main principles}\label{SectRPIgen}

The starting point for constructing the method of restricted path integrals is the Feynman formulation of quantum mechanics based on the formalism of path integrals. The Feynman path integrals are convenient for developing the main principles of the approach, being an excellent tool for conceptual analysis. Having formulated the basics of the theory, however, it will be easy to switch to the mathematically more simple formalism of Schr\"odinger equation with 
complex Hamiltonian. Schr\"odinger equation can then be used for practical calculations, without using the path integrals.

In Feynman's formulation of quantum mechanics, the amplitude of transition of the system from one point of the configuration space to another (propagator of the system) is represented by the path integral
\be
U_T(q'',q')=\int d[q]\,e^{\frac{i}{\hbar}S[q]}
=\int d[p]d[q]\,
e^{\frac{i}{\hbar}\int_0^T (p\dot q - H(p,q,t))}.
\label{Fein-int}\ee
Here $S[q]$ is the classical action of the system in question, which may be expressed as the integral of the Lagrangian along the path,
\be
S[q] = \int_0^T dt\, L(q, \dot q, t),
\label{action}\ee
and $H$ is its Hamiltonian. The first integral in \Eq{Fein-int} is the path integral in the configuration space (which may be multidimensional), and the second is the path integral in the corresponding phase space. The integrals are equal to each other, and any of them can be used for describing a closed system. For describing a continuously measured system, however, one generally needs the path integral in the phase space.

Operator $U_t$ with kernel \eq{Fein-int} --- that is,  one whose matrix elements are
\be
\la q''|U_t|q'\ra = U_t(q'',q'),
\label{U-matrix}\ee
--- is the evolution operator, and describes evolution of the system in accordance with equations
\be
|\psi_t\rangle = U_t \, |\psi_0\rangle, \quad
\rho' = U_t \, \rho_0 \, U_t^{\dagger}
\label{Unitary-evolut}\ee

We know that vector $|\psi_t\ra$ can be found by solving the Schr\"odinger equation with the initial condition $|\psi_0\ra$. The classical Hamiltonian of the system is real, and the corresponding quantum operator $H$ is Hermitian. Therefore, the evolution operator $U_t$ is unitary,
\be
{U_t}^{\dagger}\,U_t = {\bf 1}, 
\label{unitarity}\ee
and the vector $|\psi_t\ra$ has the unit norm (provided, of course, that the initial vector $|\psi_0\ra$ is normalized to unity).

To go over to the description of continuous measurement, we must first recall the conceptual basis of Feynman's representation \eq{Fein-int} of the propagator. According to Feynman's ideology, the exponential under the path integral is the amplitude of probability of transition of the system from the starting point to the end point along the given path (which may be a path in the configuration space or phase space). Since the path the transition takes is not known, the total amplitude of transition 
probability is found by carrying out summation (integration) over all possible paths, which leads to expression~\eq{Fein-int}.

The latter argument only holds, however, if it is really not possible in principle to find out along which path the system propagates. Such is indeed the case when the system is closed. The situation is different if the system is open --- that is, if the system interacts with the environment in some way or other. In this case the state of the environment is modified by the interaction, and this change will depend on the state of the system. Monitoring the change of the state of the 
environment, one can gain certain information about the evolution of the system that has caused this change. In particular, some knowledge can be obtained about the path of propagation of the system. In such case the path integral must be restricted to those paths which comply with the information obtained.

This is how the restricted path integral (RPI) arises. The idea of its application to continuous measurements was briefly formulated by Feynman in his original paper \cite{Feynman48}. This approach was developed technically and conceptually in the author's papers \cite{Men79a,Men79b,Men-bk83eng,Men-bk93} (see also Refs~\cite{AharonVardi80,Khal81RPIeng,BarchLanzProsp82,Caves86}). In two directions it has been possible to advance Feynman's ideas much further: (1) the RPI formalism has been extended to arbitrary continuous measurements, not limited to the 
continuous monitoring of the coordinates of the system; and (2) it has been proved that the calculation of RPI not only gives the probability distribution of different results of measurement, but also presents the evolution of the measured system (which in turn allows refining the first conclusions concerning the probability distribution).

\begin{figure}
\let\picnaturalsize=N
\def\picsize{3in}
\def\picfilename{corridor.eps}
\ifx\nopictures Y\else{\ifx\epsfloaded Y\else\input epsf \fi
\let\epsfloaded=Y
\centerline{\ifx\picnaturalsize N\epsfxsize \picsize\fi \epsfbox{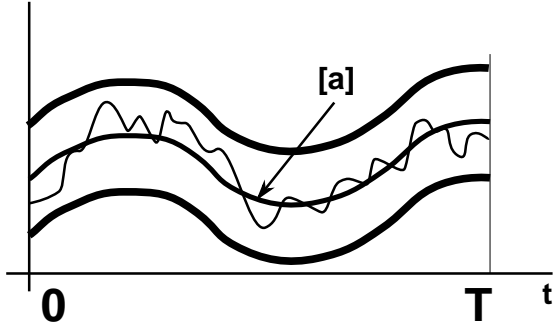}}}\fi
\vspace{5mm}
\caption{Corridor restricting the path integration in case of continuous monitoring of coordinate.
}
\label{corridor}\end{figure}

Now the easiest way to proceed is to use the most simple example of continuous measurement. Assume that the measurement consists in monitoring the position of the system in the configuration space (for definiteness, we may speak of monitoring the coordinate of a moving particle). Then at any time $t$ the measurement gives an estimate of the position --- that is, point $a(t)$ in the configuration space. In total, the result of the measurement is represented by the curve $[a]=\{ a(t) \, | \, 0\le t \le T \}$. 
Since the accuracy of measurement cannot be infinite, the resulting curve $[a]$ only gives limited information about the position of the system at any time. Namely, it tells us that at the time $t$ the system occurred in a certain neighborhood of point $a(t)$. The size of this neighborhood depends on the error of measurement $\Da$. Accordingly, the result of continuous measurement as a whole defines 
a corridor in the configuration space, centered around the curve $[a]$ and having the width of $\Da$ (see Fig.~\ref{corridor}). In view of this, the path integral in the calculation of propagator ought to be restricted to this corridor.

In more general terms, if the readout of continuous measurement $\al$ implies that path $[q]$ belongs to the family of paths $I_\al$, then the propagator of the system must be calculated by the formula
\be
U_T^{\alpha}(q'',q')
=\int_{{\bf I_{\alpha}}} d [q]\,e^{\frac{i}{\hbar}S[q]}.
\label{Ial}\ee
Similarly, if the readout  $\al$ of continuous measurement implies that path $[p,q]$ in the phase space belongs to the family of paths $J_\al$ in the phase space, then the expression for the propagator of the system is
\be
U_T^{\alpha}(q'',q')
=\int_{{\bf J_{\alpha}}} d[p]d[q]\,
e^{\frac{i}{\hbar}\int_0^T (p\dot q - H(p,q,t))}.
\label{Jal}\ee
Thus, we find the propagator as the integral over the corridor of paths. By analogy with Carmichael's quantum trajectory \cite{Carmichael-bk93}, we may refer to $I_\al$ or $J_\al$ as the {\em quantum corridor}.

The corridor in the configuration space $I_\al$ may be regarded as a particular case of the corridor $J_\al$ in the phase space. This special case is realized when the information gained from the measurement is adequately expressed in terms of coordinates and does not require the involvement of momentum. Such is the case, for example, when the measurement consists in monitoring the coordinate.\footnote{Notice that the information about the path $[q]$ obtained from the measurement allows making certain conclusions concerning the velocity $[\dot q]$, but this does not yet automatically give the momentum, since the classical relation $p=m\dot q$ does not hold in the quantum regime (see Refs~\cite{Men92moment,Men-bk93}).}
Usually the information supplied by the measurement does not allow defining the corridor of paths with clear-cut limits. A more adequate description of the measurement is based on the weight functional
\be
U_T^{\alpha}(q'',q')
=\int d [q]\,w_\alpha[q]\,e^{\frac{i}{\hbar}S[q]}
\label{Int-wq}\ee
or, in the general case,
\be
U_T^{\alpha}(q'',q')
=\int d[p]d[q]\,w_\alpha[p,q]\,
e^{\frac{i}{\hbar}\int_0^T (p\dot q - H(p,q,t))}. 
\label{Int-wp}\ee

In such situation we may also speak of a quantum corridor, but the boundaries of this corridor are blurred. If, for example, measured continuously is the coordinate $q$, and the measurement readout is represented by curve $\al=[a]$, then the functional $w_\alpha[q]$ must be equal to one for paths $[q]$ lying close to curve $[a]$, and turn into zero for paths straying far from this curve.

\begin{remark}\label{remarkComplexWeight}
{\rm A still more realistic description of the measurement requires using a complex-valued functional $w_\alpha[p,q]$ in place of the real-valued one. This means that the measurement involves not only projecting the system on to the subspace corresponding to the measurement readout, but also a change of phase of the wave function depending on the readout. Such measurement is not minimal 
because the same information can be obtained without distortion of phase (while the 
projection is absolutely necessary). Real devices, however, may realize exactly such nonminimal measurements. Besides, for some purposes a nonminimal measurement may be advantageous.}
\end{remark}

Having expressed in one way or another the propagator $U_T^{\alpha}(q'',q')$ and passing to the relevant evolution operator $U_T^{\alpha}$ using the formula similar to \Eq{U-matrix}, we get the law of evolution in the form (provided that the continuous measurement in question is performed and yields the desired result)
\be
|\psi_T^{\alpha}\rangle = U_T^{\alpha} |\psi_0\rangle, \quad
\rho_T^{\alpha} = U_T^{\alpha} \rho_0 \left(
U_T^{\alpha}\right)^{\dagger}
\label{evolut-meas}\ee
Now we have a whole family of {\em partial evolution operators} $U_T^{\alpha}$, and these operators are not unitary. Vector $|\psi_T^{\alpha}\rangle$ produced by the application of such operator has the norm less than one (even if the initial vector is normalized). The trace of the density matrix $\rho_T^{\alpha}$ is less than one, even if the initial density matrix has a unit trace. This circumstance is not accidental, because the new norms give us the probability distribution of different measurement readouts. Namely, the quantity
\be
P(\al)=\tr\rho_T^{\alpha}
=\tr\left( U_T^{\alpha}
\rho_0 \left( U_T^{\alpha}\right)^{\dagger}\right)
\label{Prob-al}\ee
is the density of the probability that the measurement will yield the result $\alpha$. The distribution is normalized with respect to a certain measure $d\alpha$, so that
\be
\int d\al\, P(\al) = 1.
\label{Prob-norm}\ee

Carrying out summation (using this measure) over all possible outcomes of measurement $\al$, we may go over to the nonselective description of continuous measurement:
\be
\rho_T = \int d\alpha\,  \rho_T^{\alpha}
=\int d\alpha\, U_T^{\alpha} \rho_0 \left( U_T^{\alpha}\right)^{\dagger}.
\label{non-select-ro}\ee
Then the measurement readout is assumed to be unknown, and the evolution of the measured system is described by the density matrix. The trace of the density matrix is unity for any initial state $\rho_T$ of the system, as long as the condition of generalized unitarity
\be
\int d\alpha\,  \left( U_T^{\alpha}\right)^{\dagger}\, U_T^{\alpha} =
\bf 1 
\label{genUnitar}\ee
is satisfied. This condition ensures preservation of probability (30), and therefore is mandatory for the family of partial evolution operators.

\subsubsection{Monitoring of arbitrary observable}

Embarking on the construction of the RPI formalism, we selected the monitoring of coordinate as our working example of continuous measurement. Let us now consider a more general case of monitoring of an arbitrary observable which may be a function of coordinates, momenta and time, $A=A(p,q,t)$. The result of monitoring of this observable is represented by the function
\be
[a] = \{a(t) \, | \, 0\le t \le T\}. 
\label{res-a}\ee
This result implies that the observable $A$ at any time $t$ is close to $a(t)$. In other words, curve $[A]$, as determined by its values
\be
A(t)=A(p(t),q(t),t),
\label{Aoft}\ee

is generally close to curve $[a]$. The word ``close" is rather loose and ought to be given a definition. For the criterion of closeness of two curves we shall select the mean square distance between them,
\be
\langle (A-a)^2\rangle_T
= \frac{1}{T}\int_0^T
[ A(t) - a(t) ]^2\,dt.
\label{quadr-deflect}\ee

Now we define the weight functional in \Eq{Int-wp} as
\be
w_{[a]}[p,q] 
= \exp\left(-\kappa \langle (A-a)^2\rangle_T \right)
= \exp\left(
{-\kappa \int_0^T [ A(t) - a(t) ]^2\,dt }
\right).
\label{gauss-weight}\ee
Thus we have in a certain way concretized the concept of the quantum corridor which describes our measurement (monitoring of the observable $A$). We may say that for the description of monitoring we are using Gaussian corridor.

The measure of closeness of the two curves and the weight functional can be chosen in different ways, and the resulting description of the measurement will to a certain extent depend on this choice. Having decided in favor of a particular choice, we solidify the theory. Selecting different options, we shall be getting somewhat different description for the measurement which we have called the monitoring of observable $A$. This uncertainty has a straightforward 
physical interpretation --- it is related to the freedom of physical realization of the monitoring.

Mathematically, our definition of the weight functional \eq{gauss-weight} is the simplest. At the same time, it can be proved that this functional correctly describes the behavior of the system when the monitoring is realized in the form of frequent and short (almost instantaneous) observations of the system --- that is, as a series of weak interactions with the measuring system. In principle, however, one can select a different weight functional, thus describing a different class of measuring devices (see Remark~\ref{remarkComplexWeight} above).

Coefficient $\kappa$ in \Eq{gauss-weight} characterizes the ``force of measurement", its accuracy. To better understand the physical meaning of this parameter, we may represent it in the form
\be
\kappa=\frac{1}{T\Delta a_T^2}.
\label{kappa-Da}\ee
Substituting this expression for $\kappa$ into Eqn (36), we see that $\Da_T$ is the error of the continuous measurement that lasts for the time $T$. Indeed, the weight functional (36) is designed so that a substantial contribution into the RPI only comes from those paths which correspond to functions $A(t)$ whose mean square deviation from $a(t)$ is not greater than $\Da_T$.

If the parameter $\kappa$ remains constant with the time, then $\Da_T$ falls,
\be
\Delta a_T^2\sim \frac{1}{T}, 
\label{Da-T}\ee

that is, the resolution of continuous measurement improves as its duration increases. We shall return to this effect later on, in connection with the continuous measurement of energy. At this point we just note that it leads to the exponentially fast decoherence of the type of \Eq{exp-time-deco}.

Using the Gaussian definition \eq{gauss-weight} of quantum corridor, we may rewrite the restricted path integral \eq{Int-wp} in the form
\begin{eqnarray}
U_T^{[a]}(q'',q')&=&\int d[p]\,d[q]\,
\exp\left\{ \frac{i}{\hbar} \int_0^T \big(p\dot q
- H(p,q,t)\big)\,dt \right.\nonumber\\
&-& \left.\kappa \int_0^T \big(A(p,q,t)-a(t)\big)^2 dt
\right\}.
\label{gaussRPI}\end{eqnarray}

This integral will coincide with the conventional (nonrestricted) Feynman integral \eq{Fein-int} if we replace the Hamiltonian $H$ by the effective complex Hamiltonian
\be
H_{[a]}\,(p,q,t) = H(p,q,t) - i\kappa\hbar \,\big( A(p,q,t) - a(t)
\big)^2.
\label{EffHamilt}\ee

This means that the evolution described by the partial evolution operator \eq{evolut-meas} can be equivalently described by Schr\"odinger equation with the Hamiltonian \eq{EffHamilt}:
\be 
|\dot\psi\ra 
=  -\frac{i}{\hbar} H_{[a]}|\psi\ra
= \left[ -\frac{i}{\hbar} H 
- \kappa\big( A-a(t)\big)^2 \right] |\psi\ra. 
\label{SchroedEffEq}\ee 

This equation describes evolution of the measured system in the selective way --- that is, with due account for the readout of measurement $[a]$. If we go over to the nonselective description, carrying out integration over all $[a]$ in accordance with \Eq{non-select-ro}, then the resulting density matrix will satisfy \cite{Men94MasterEq} the Lindblad equation~\eq{LindbladEq}.

\begin{remark}
{\rm The arguments developed above hold if the time in the course of monitoring of the observable $A$ is measured with absolute precision (that is, if the finite accuracy of timing can be neglected). Of course, in reality the quantity $a(t)$, derived in the course of monitoring, characterizes the value of the observable on a certain time interval, whose length characterizes the accuracy of measurement of time, or, looking from a different angle, the inertial properties 
of the measuring device. The general scheme outlined in Section~\ref{SectRPIgen} was extended to this case in Ref.~\cite{Men97timeResol}. It was demonstrated that, in case of finite time resolution, evolution of the system cannot be described by a Schr\"odinger-type differential equation, and the measurement is ``time-integral". The RPI technique of representation of propagators remains workable in this case as well.
}
\end{remark}

\subsubsection{*Uncertainties in continuous measurements}

A continuous measurement is completely described by the set of partial propagators \eq{Int-wp}, each of which can be found by calculating the RPI or by solving the Schr\"odinger equation with complex Hamiltonian. This, however, is not quite easy to do. Then a natural question is whether it might be possible to use simpler procedures (like the uncertainty relations) for solving at least some of the problems related to the continuous measurement. This indeed can be done. Namely, there are 
rather simple methods that give an approximate answer to the question as to which measurement readouts $[a]$ occur with a high enough probability.

This answer is expressed by inequality which is called the principle of uncertainty of action 
\cite{Men91AUP}-\cite{Men96AUP}, 
since it contains the estimate of how the classical action is changed when the path is varied within one and the same quantum corridor $[a]$.

The condition that the probability of the measurement readout $[a]$ be high is formulated in terms of one of the paths $[p,q]$ reconciled with this readout --- that is, satisfying the condition $A(t)=a(t)$, where $A(t)=A(p(t),q(t),t)$.\footnote{It can be demonstrated that the proof does not depend on which path $[p,q]$ reconciled with the measurement readout $[a]$ is selected.} Apart from this perfectly matching path, we consider a path $[p+\D p,q+\D q]$ which differs from the latter by the increment $[\D p, \D q]$ (where $\D p=\D p(t)$, $\D q=\D q(t)$ are functions of time). Then the function $A(t)$ has the increment of $\D A(t)$. We only consider increments of the path such that the increment $\D A(t)$ remains less 
than the error of the measurement $\D a(t)$. This means that the new path $[p+\D p,q+\D q]$ within the accuracy of measurement $\D a$ still agrees with the measurement readout $[a]$. To simplify the terminology, we shall say that the new path lies within the corridor $[a]$ --- that is, the increment of the path does not lead beyond this corridor.

Now we find the classical action for the path $[p,q]$ by the formula
\be
S[p,q] = \int_0^T (p\dot q - H(p,q,t))
\label{Spq}\ee
and look at the increment $\D S$ upon transition to the new path $[p+\D p,q+\D q]$. The magnitude of the increments of action found in this way is the criterion of how probable is the measurement readout $[a]$. It turns out that the corridor $[a]$ is probable if the increment of action $\D S$ does not exceed the quantum of action $\hbar$ until the path lies within the corridor $[a]$.\footnote{The background of this criterion is obvious. If variation of the path within the corridor results in large variations of action, the imaginary exponential of action in the integral \eq{gaussRPI} exhibits fast oscillations, and the integral is small. Since this integral defines the probability density for the measurement readout $[a]$, the probability is also small.}

This condition ensures that the measurement readout $[a]$ occurs with a high enough probability. It can be written in the following compact form:
\be
\max_{|\D A(t)|\le\Da(t)}\left|\sum_i
\int_0^T dt\,\left[
\D p_i\left(\dot q_i - \partderiv{H}{p_i}\right)
- \D q_i\left(\dot p_i + \partderiv{H}{q_i}\right). 
\right]\right|\lo\hbar
\label{AUP-2}\ee
The parentheses under the integral enclose expressions that enter the Hamilton equations; accordingly, this inequality can be interpreted in terms of ``fictitious force". Namely, the measurement readout $[a]$ is probable if the corresponding path $[p,q]$ satisfies the modified Hamilton equations
\ba
\dot q_i - \partderiv{H}{p_i}
&=& - \dF (t) \partderiv{A(p,q,t)}{p_i} \\
\dot p_i + \partderiv{H}{q_i}
&=& \dF (t) \partderiv{A(p,q,t)}{q_i}
\label{fictForce}\ea
with the ``fictitious force" $\dF (t)$ subject to the restriction
\be
\int_0^T dt\, |\dF (t)|\,\D a(t)\lo\hbar. 
\label{fictRestrict}\ee

\begin{figure}
\let\picnaturalsize=N
\def\picsize{4.0in}
\def\picfilename{uncert.eps}
\ifx\nopictures Y\else{\ifx\epsfloaded Y\else\input epsf \fi
\let\epsfloaded=Y
\centerline{\ifx\picnaturalsize N\epsfxsize \picsize\fi \epsfbox{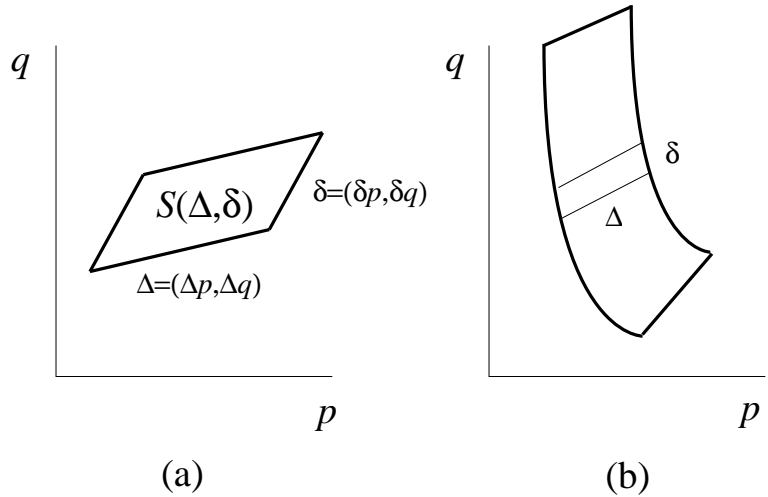}}}\fi
\caption{(a) Area of parallelogram in phase space built on two vectors. (b) Area in phase space defined by two paths $[p,q]$, $[p+\D p,q+\D q]$, lying within the corridor $[a]$. If this area is less than $\hbar$ for each such pair of paths, then $[a]$ will occur as the measurement readout with a sufficiently high probability.} \label{figUncert}\end{figure}

Alternatively, the condition \eq{AUP-2} can be written as the restriction on the area in the phase space defined by the two paths $[p,q]$ and $[p+\D p,q+\D q]$ (see Fig.~\ref{figUncert}):
\be
\max_{|\D A(t)|\le\Da(t)}\left|
\sum_i \int_0^T dt\,\delta\sigma_i \right| \lo\hbar
\label{phase-area-restr}\ee
where
\ba
\delta\sigma_i(t)=\D p_i(t)\,\delta q_i(t)
- \D q_i(t)\,\delta p_i(t)\nonumber\\
\delta q_i = dq_i - \frac1m p_i dt, \quad
\delta p_i = dp_i - F_i dt. 
\label{phase-area}\ea

Inequality \eq{phase-area-restr} must hold for all paths that belong to this corridor. This implies that for some of the paths the equality (by order of magnitude) will be attained. For a one-dimensional system we then have
\be
|\int (\D p \,\delta q - \D q \,\delta p)| \sim \hbar. 
\label{phase-area-restr1}\ee
If the measurement is very short, we may write
\be
|(\D p \,\delta q - \D q \,\delta p)| \sim \hbar
\label{phase-area-restr1sh}\ee

Consider the measurement of momentum with the accuracy of $\D p$ over the time interval $\D t$. If the coordinate is not measured, then $\D q$ is very large, and at first sight it may seem that the second term in \Eq{phase-area-restr1sh} dominates. In this case, however, $\delta p$ must be exactly equal to zero, since otherwise the inequality \eq{phase-area-restr} would not be satisfied. Ergo, the second term vanishes, and we have
\be
\D p \,|\delta q|
=\D p \left|(q''-q') - \frac pm \,\D t \right| \sim \hbar 
\label{measMoment}\ee
where $q'$, $q''$ are the coordinates before and after the measurement. We see that the effect of measurement may consist in that the velocity $(q''-q')/\D t$ differs from the classical expression $p/m$ by the amount of the order of $\hbar/\D p\D t$. This is an alternative formulation of the well-known uncertainty relation \cite{LandauLifsh87eng}
\be
\D p \, |v''-v'| \D t \sim \hbar, 
\label{uncertDpDv}\ee
which is deduced from thought experiment. Relation \eq{phase-area-restr}, however, is much more general.

\subsubsection{Features of RPI-based approach}

Let us discuss some important features of the approach based on restricted path integral. We saw that this approach is deduced from first principles --- namely, from quantum mechanics in Feynman's formulation. At the same time, it can be given an independent substantiation based on the concrete models of measurement (see Section~\ref{SectRealization} below). Validation of this phenomenological approach with models makes it more reliable, while the derivability from first principles points to the fundamental character of the theory.

According to the RPI-based approach, the continuously measured system is described as an open system, but the description is selective. Evolution of the measured (open) system is described by a state vector rather than by a density matrix. Such description is applicable to an individual system, and not only to the statistical ensemble of 
systems corresponding to all possible states of the measuring medium. The RPI approach reveals a very important feature of the interaction between the measured system and the measuring system: the reciprocal effect of the measurement on the measured system depends only on the information obtained from the measurement (and recorded in the state of the environment). In this sense the RPI approach to continuous measurements may be called the information approach \cite{Men97rev}.

The dynamics of the measured system, described by the set of partial propagators \eq{gaussRPI} or by the Schr\"odinger equation with the effective complex Hamiltonian \eq{SchroedEffEq}, depends only on the information recorded by the measuring medium, but not on the particulars of the interaction of the system with the environment. This is a manifestation of the dynamic role of information. The complete description of the open (measured) system does not require a complete model of the environment. 
The information model is quite sufficient.

As a matter of fact, the dynamic role of information is already clear from von Neumann's reduction postulate, which holds that the final state of the measured system only depends on the outcome of the measurement. In the reduction postulate, however, the change of the system caused by the measurement and the evolution of the system owing to its own dynamic properties are time-separated (cf. \Eq{serialMeas}).

By contrast, the two dynamic aspects are inseparable in the evolution of the continuously measured system. Such systems exhibit a new type of dynamics which involves both classical and quantum features at the same time. The dynamics is completely determined by fixing (1)~the Hamiltonian of the system, and (2)~the information about the system that flows (dissipates) into the environment.

The RPI approach proves that the theory of measurements can be incorporated into quantum mechanics, contrary to the common opinion that it ought to be postulated independently of quantum mechanics. One only has to understand quantum mechanics the way it has been formulated by Feynman. This indicates that Feynman's formulation of quantum mechanics is essentially broader and deeper than its conventional operator form. The added depth 
arises from the fact that physical interpretation is given not only to the full amplitude of propagation of the system, but also to the amplitude of propagation along the given path. This formulation of quantum mechanics is closed because it naturally incorporates the quantum theory of measurements.

In Section~\ref{MeasEn} we shall discuss in detail an important application of the RPI method to the continuous measurement of the energy of a multilevel system. At this point we shall just mention some other applications of this method in nonrelativistic and relativistic quantum theories.

In the nonrelativistic theory this method has been applied to the analysis of the measurement of the coordinates of oscillator and a system of oscillators \cite{Men79a,Men79b,Men-bk83eng}, and to quantum nondemolition measurements \cite{GolubMen89qnd,Men96AUP,MenAud97qnd}. The relativistic applications include:
\begin{itemize} 
\item quantum restrictions on the measurability of electromagnetic field~\cite{BorzMen94electromagn},

\item quantum restrictions on the measurability of gravitational field~\cite{Men85measGrav},

\item appearance of classical geometry in quantum gravitation~\cite{Men91quCosm},

\item measurement of the position of relativistic particle~\cite{MenBorz95posit},

\item analysis of Unruh and Hawking thermal effects~\cite{Men98unruhEng}.
\end{itemize} 

\subsection{*Derivation of the stochastic equation}
\label{Sect-deriv-stochast}

The Schr\"odinger equation with complex Hamiltonian \eq{SchroedEffEq} can be written as the stochastic Schr\"odinger equation \cite{PreOnoTam96}. To do this, in place of the function $a(t)$ and the state vector $\psi(t)$ we introduce new variables
\be
a=c+\frac{\xi}{\sqrt{2\k}}, \quad
\Psi(t)=\exp\left(
\frac 12 \int_0^t dt\, \xi^2
\right)\psi(0),
\label{ksiPsi}\ee
where $c(t)$ will be defined later. Then \Eq{SchroedEffEq} can be rewritten as
\be
|\dot\Psi\ra=\left[ 
-\ioverh H -\k(A-c)^2 + \sqrt{2\k} (A-c)\xi
\right]|\Psi\ra
\label{EqPsi}\ee
or, if we define a new variable $w$ as $dw=\xi dt$,
\be
d|\Psi\ra=\left[ 
-\ioverh H -\k(A-c)^2 \right] |\Psi\ra dt
+ \sqrt{2\k} (A-c) |\Psi\ra dw.
\label{StochEq2}\ee
In order to have the norm of vector $|\Psi\ra$ conserved,
\be
(\la \Psi| + \la d\Psi|)(|\Psi\ra + |d\Psi\ra)= \la\Psi | \Psi\ra, 
\label{normPsi}\ee
it is sufficient to require that
\be
dw^2=dt, \quad c=\la\Psi |A| \Psi\ra. 
\label{stochCondit}\ee

Equation \eq{StochEq2} with the additional conditions \eq{stochCondit} is known as the stochastic Schr\"odinger equation. The quantity $w$ here is a random function (white noise), which describes the impact of the measuring medium on the measured system. The statistics of this noise is determined by the probability distribution
\be
P[w] = \la\psi |\psi\ra
=\exp\left( -\frac 12 \int_0^t \dot w^2(t) dt\right).
\label{Prob-w}\ee
This is the so-called white noise or Brownian motion. In case of such motion, the change of $w$ over a small time interval $\Delta t$ is described by the distribution
\be
P(\D w)=\frac{1}{\sqrt{2\pi\D t}}
\exp\left( -\frac 12\frac{\D w^2}{\D t}\right), 
\label{Brownian}\ee
and, as a consequence, the mean square displacement is proportional to the time:
\be
\overline{\D w^2} 
= \int P(\D w) \D w^2 d\D w
= \D t. 
\label{Dw2}\ee
This last formula offers a physical interpretation of the unusual relation $dw^2=dt$, which is adopted in the stochastic theory in the framework of the so-called Ito's formalism~\cite{Ito}.

In this way, the stochastic equation \eq{StochEq2} is derived from the theory of continuous measurements based on the restricted path integral or on the Schr\"odinger equation with complex Hamiltonian. Other stochastic equations have also been proposed for describing continuous quantum measurements \cite{Gisin89stochast,Diosi89stochast,Belavkin89stochast}. The necessary condition is that the density matrix, which describes the same system in the nonselective manner, satisfy the Lindblad equation \eq{LindbladEq}. This 
holds for Eqn \eq{StochEq2} because the Lindblad equation is an implication of the RPI approach (see~\cite{Men94MasterEq}).

\subsection{*Consistent histories}

A new direction of research in quantum mechanics, started a few years ago, is based on the concept of consistent histories 
\cite{Griffiths84}-\cite{GellMannHartle93}. 
This research is concerned with the emergence of classical features in quantum systems --- that is, the effect of decoherence.

The approach based on consistent histories bears some technical semblance to the RPI method, although it is radically different. The difference is that consistent histories are used for describing a closed quantum system (with the purpose of detecting <<their>> classical features), whereas the RPI method deals from the start with an open system, while the effects of the environment are taken into account implicitly. Let us briefly discuss the method of consistent histories, following the 
paper by Gell-Mann and Hartle~\cite{GellMannHartle90}.

A history $\al=\{ i_1, i_2, \dots , i_{N-1} \}$ is defined in Ref.~\cite{GellMannHartle90} as a chain of projectors that specify the state of the system at successive times. These projectors define the evolution operator
\be
U_{\al}=
U(t_N,t_{N-1})P_{i_{N-1}}U(t_{N-1},t_{N-2})\dots 
U(t_3,t_2)P_{i_2} U(t_2,t_1)P_{i_1} U(t_1,t_0)
\label{consistUal}\ee
which is similar to that used in \Eq{serialMeas} for describing a series of instantaneous measurements. In the method of consistent histories, however, the projectors are used not for describing real measurements, but rather for analyzing the free evolution of the system without any external influence. The operator $U_{\al}$ is just one quantum alternative out of many that contribute to the evolution of the quantum system. The total evolution operator is the sum
\be
U = \sum_{\al}U_{\al} 
\label{consistU}\ee
over all possible histories --- that is, over all possible selections of projector for each moment of time\footnote{By contrast, if operator \eq{consistUal} describes a real measurement, as has been assumed in \Eq{serialMeas}, it corresponds to the classical alternative, and summation of such operators is meaningless.}

Histories in the sense of Gell-Mann and Hartle are straightforward analogs of Feynman's paths, which determine the evolution of a closed system only when taken together. By contrast to individual paths, histories are more coarse-grained alternatives, each of which includes many paths.

According to Gell-Mann and Hartle, each history is associated with the ``probability"
\be
P_\al=\tr (U_\al \rho U_\al^\dagger).  
\label{consistPal}\ee
The interpretation of this quantity as a probability is far not necessarily correct, and further analysis is concerned with the question when exactly such interpretation is justified.

Along with a certain selected set of histories $\{\al\}$, the authors also consider sets of histories which follow from the selected set by applying the procedure of coarse-graining --- when the projections are made onto more extensive subspaces, and/or are less frequent (not at every selected moment of time). Each history from the coarser set can be represented as a sum of histories from the finer set (a coarse history includes fine histories as its alternatives). The corresponding evolution operators are bound by the summation
\be
\beta=\sum_{\al\in\beta}\al, \quad 
U_\beta=\sum_{\al\in\beta} U_\al. 
\label{CoarseSum}\ee
At the same time, the corresponding ``probabilities" of the coarser histories
\be
P_\beta=\tr (U_\beta \rho U_\beta^\dagger)
\label{consistPbeta}\ee
are not necessarily representable as sums of ``probabilities" of the finer histories.

The key to the entire approach is found in the following argument which leads to the ``consistency condition" as the necessary criterion that the description of the system in terms of alternatives $\{\al\}$ is classical. A sufficiently coarse-grained description of quantum system loses its quantum features and becomes purely classical. The alternatives $\{\al\}$ give a coarse-grained description of the system, and if the description is coarse enough, these 
alternatives may be regarded as classical. In this case, the quantities \eq{consistPal} indeed ought to be considered as probabilities complying with the conventional rule of summation.

Now if the alternatives from the set $\{\al\}$ can be regarded as classical alternatives, then any coarser set of alternatives $\{\beta\}$ is also classical, and therefore the {\em rule of summation of probabilities} holds:
\be
P_\beta=\sum_{\al\in\beta} P_\al.  
\label{consistProbSum}\ee

Thus, if the alternatives $\{\al\}$ are classical, then the transition to the coarser alternatives $\{\beta\}$ complies not only with the rule of summation of amplitudes \eq{CoarseSum}, but also with the rule of summation of probabilities \eq{consistProbSum}. A straightforward mathematical analysis proves that condition \eq{consistProbSum} is satisfied if the functional of decoherence
\be
C_{\al\al'}=\tr (U_\al \rho U_{\al'}^\dagger)
\label{decoherFunct}\ee
satisfies the consistency condition
\be
C_{\al\al'}+C_{\al\al'}^{*}=0 \quad \mbox{for} \quad \al\ne\al'.
\label{consistCondit}\ee
We see that the consistency condition \eq{consistCondit} is necessary for the set histories $\{\al\}$ to be interpreted as a set of classical alternatives.

It was demonstrated that the condition of consistency holds approximately for the ``situation of measurement" --- that is, when the system under consideration consists of two parts which interact like the measuring device and the measured subsystem (see Section~\ref{Sect-superselect}). At the same time, the condition of consistency of histories is not sufficient for ensuring the predictability of the quantum system --- a criterion providing that the 
description is classical indeed \cite{PazZurek93}. We see that the theory of consistent histories, certainly being a step forward in the study of decoherence, does not yet give a complete solution of the problem.

Drawing comparison once again between the two approaches, the methods of RPI and consistent histories, we duly note that the condition of consistency is not at all necessary in the RPI approach. This is because the RPI approach is 
concerned with an open system that interacts with its environment (although there is no explicit model of this environment). The RPI-based analysis of the system involves only one family of histories $\{\al\}$ --- the family that describes the effects of the environment. There is no need of using the coarser sets of histories, and of providing consistency of descriptions at different levels of coarse-graining.

\section{Continuous measurement of discrete energy}
\label{MeasEn}

An important achievement of recent years is the analysis of continuous energy measurement in a system with discrete energy levels. Firstly, the measurement of the energy of atom is of much practical importance. Secondly, all important features of the continuous quantum measurement in this case can be transferred to continuous measurement of any discrete observable. We are going to consider the continuous energy measurement in a multilevel (especially a two-level) system, and 
prove that this process allows monitoring a quantum transition from one level to another. First we shall analyze the measurement using the Schr\"odinger equation with complex Hamiltonian, and then consider a concrete scheme of such measurement using conventional quantum mechanical methods. We shall also show the way of implementing this measurement.

The approach based on RPI and complex Hamiltonians was first applied to the measurement of energy in a two-level system in Refs~\cite{OnoPreTam93En,OnoPreTam95En}. It was demonstrated that if the energy is measured with a high enough accuracy, then the system becomes frozen, and the transitions between the levels are no longer possible (the Zeno effect). The alternative regimes of measurement have not been (and could not have been) studied in these works because of a serious methodological error. 
The authors assumed that the result of a continuous measurement is expressed by the function $E(t)$, which does not change and coincides with one of the energy levels of the system.

At first sight, such assumption seems natural for a system with discrete spectrum. However, it is not correct, since the accuracy of the measurement is finite. Function $[E]$ which represents the measurement readout can be arbitrary in the approach based on RPI or on the Schr\"odinger equation with complex Hamiltonian. By solving the Schr\"odinger equation (similar to \Eq{SchroedEffEq}) with the $[E]$-dependent Hamiltonian, one can 
find the probability density $P[E]$ of the given result of measurement. It is only then (and by no means a priori) that one can decide which functions $[E]$ may arise as the measurement readout. It turns out that in the regime of a very accurate measurement, probable are only those results that correspond to functions $[E]$ which 
are constant and coincide with the energy levels. This is not true, however, when the accuracy of the measurement is relatively poor. But it is exactly those imprecise or soft measurements that are of special interest, since they do not modify the measured system too much.

This error was corrected in Ref.~\cite{AudMen97En}, which has made it possible to carry out a detailed analysis of a moderately accurate continuous quantum measurement of the energy, and propose an entirely new type of measurement --- monitoring of a quantum transition. The results that could be anticipated from such measurement were thoroughly analyzed in Ref.~\cite{AuMenNam97scat}, and a concrete scheme of continuous quantum measurement of the energy of the atom was proposed.

\subsection{Energy measurement in a multilevel system}
\label{MeasEnMultilevel}

Consider a system with the Hamiltonian $H=H_0+V$, where $H_0$ is the ``free" Hamiltonian of the multilevel system, and $V$ is the perturbation that may induce a transition between the levels. Assume that the observable $H_0$ is being continuously measured in this system. In accordance with the scheme outlined in Section~\ref{RPI}, we shall describe such measurement with the functional
\be
w_{[E]}[p,q] = \exp\left( -\kappa \int_0^T
[ H_0\big(p(t),q(t),t\big) - E(t) ]^2\,dt \right),
\label{weight-En}\ee
that is, with the Gaussian corridor centered around curve $[E]$. Then the effective Hamiltonian is
\be
H_{[E]} = H_0  + V - i\kappa\hbar \,\big( H_0 - E(t) \big)^2
\label{EffHam-En}\ee
and the effective Schr\"odinger equation with the complex Hamiltonian is
\be
\frac{\partial}{\partial t} |\psi_t\rangle
  = \left(-\frac{i}{\hbar} H
  -\kappa \,\Big( H_0 - E(t)\Big) ^2\right)\, |\psi_t\rangle.
\label{EffEq-En}\ee

Solving this equation, we can find vector $|\psi_t\ra$ at any time. In accordance with the general formula (29), the norm of this vector at the end time of the continuous measurement gives the density of the probability that the measurement readout $[E]$ will be realized:
\be
P[E]=||\psi_T||^2. 
\label{Prob-En}\ee

Using the expansion of the state vector with respect to the basis
\be
|\varphi_n(t)\rangle = e^{-iE_n\, t/\hbar}|n\rangle,
\label{En-basis}\ee
we get the following set of equations for the expansion coefficients:
\be
\dot C_n = -\kappa (E_n-E(t))^2\,C_n
  -\frac{i}{\hbar}\sum_{n'}\langle\varphi_n|V|\varphi_{n'}\rangle C_{n'}
\label{EffEq-En-Cn}\ee
and the formula for the probability distribution of the measurement readouts
\be
P[E]=\sum_n|C_n(T)|^2. 
\label{Prob-En-Cn}\ee

For a free multilevel system, $V=0$, the solution has a simple form:
\be
C_n(T)=C_n(0)\exp \left[ - \kappa\int_0^T dt
\left( E_n - E(t)\right) ^2\right]
= C_n(0)\exp \left[
-\frac{T}{T_{\rm lr}}\,\frac{\langle (E_n-E)^2\rangle_T}{{\Delta 
E}^2}\right], 
\label{FreeSys-Cn}\ee
where
\be
T_{\rm lr}=\frac{1}{\kappa\Delta E^2}
\label{Tlr}\ee
is the timelike parameter which characterizes the accuracy of measurement. It is expressed in terms of the ``typical energy difference" $\DE$ in the part of spectrum we are concerned with.

From \Eq{FreeSys-Cn} we see that if $T\gg T_{\rm lr}$ (that is, the duration of measurement is large enough compared to the characteristic parameter $T_{\rm lr}$), then the regime of measurement is realized that resolves (separates) the energy levels. Its characteristic features are:
\begin{itemize}
\item $[E]$ is a function which is close to one of the energy levels, $E(t)\simeq E_n$;

\item the system after the measurement occurs at level $n$;

\item the probability of $[E]$ being close to $E_n$ is $|C_n(0)|^2$.
\end{itemize}
Indeed, if function $[E]$ is close (in the mean square sense, cf. \Eq{quadr-deflect}) to one of the levels $E_n$, then the coefficient $C_n$ with the relevant number  remains after the measurement (at $t=T$) the same as it was before the measurement, while all the other coefficients are exponentially small. According to \Eq{Prob-En-Cn}, the probability density of each such function is $|C_n(0)|^2$.

If function $[E]$, however, does not satisfy this condition (does not remain close to one and the same level in the course of the entire measurement), then all coefficients $C_n$ at the time $T$ are exponentially small. The probability density of each of these functions is negligibly small.

So we have to conclude that only those results of measurement are realized with a reasonable probability which are represented by functions close to one of the levels. The probability that the readout is close to $E_n$ is proportional to $|C_n(0)|^2$. From considerations of normalization we deduce that this probability is equal to $|C_n(0)|^2$ (which can be proved more rigorously). It is easy to see that these results are in perfect agreement with the way the von Neumann energy measurement is described.

If the measurement is not long enough, $T\ll T_{\rm lr}$, then the regime of measurement is not capable of resolving the energy levels:
\begin{itemize}
\item Variation $E_{\rm max}-E_{\rm min}$ of curve $[E]$ is less than $\Delta E\sqrt{T_{\rm lr}/T}$ and can be much greater than $\DE$;

\item $C_n(T)\simeq C_n(0)$ for levels between $E_{\rm min}$ and $E_{\rm max}$;

\item $C_n(T)$ are exponentially small outside of $[E_{\rm min},E_{\rm max}]$.
\end{itemize}

Thus, by the time $T$ all energy levels vanish outside of the interval of the width $\Delta E\sqrt{T_{\rm lr}/T}$. As $T$ increases, this interval narrows down, and by the time $T=\Tlr$ there is only one energy level --- that is, the system occurs in the state with the definite energy $E_n$. Curve $[E]$ by this time is very close to the level $E_n$.\footnote{Closeness is understood in the sense of mean square deviation, so the deviation can actually be very large for a very short time.} The probability that level number $n$ will survive is $|C_n(0)|^2$.

We see that when the time of measurement is not yet long enough, $T\ll T_{\rm lr}$, the energy cannot be measured with sufficient precision, and the system still occurs in the state that is a superposition of a number of levels. When the duration of continuous measurement becomes greater than $\Tlr$, the result of the continuous measurement fixes on one of the levels, and the system goes to this level. Accordingly, the time parameter $\Tlr$ may be interpreted as the {\em level resolution time}.

From arguments developed above it is clear that the continuous measurement of energy as realized in the method of quantum corridors is a model of the process of decoherence for the von Neumann ``instantaneous" measurement \cite{AudMen97En}. By this we mean the following. The von Neumann scheme assumes that the measurement of energy in a multilevel system occurs instantaneously, yields (with the relevant probability $|C_n(0)|^2$) one of the values of $E_n$, and the system after the measurement 
occurs at the $n$th level. In reality, all this takes some finite time which is neglected when the measurement is described in accordance with the von Neumann scheme. The continuous measurement reveals the time structure of this transition --- that is, describes what we call the process of decoherence.

Let us now consider a two-level system under a resonant force. Let $H_0$ be a Hamiltonian of the two-level system, and the potential $V$ defined by its matrix elements
\be
\langle\varphi_1|V\varphi_2\rangle=\langle\varphi_2|V\varphi_1\rangle^{*}=V_0.
\label{resonance-potential}\ee
If we go over from basis \eq{En-basis} to basis $|n\ra$, then the matrix elements of operator $V$ will be harmonic functions of frequency $\omega=\D E/\hbar$ --- that is, the potential $V$ will describe the resonant action. The set of equations \eq{EffEq-En-Cn} becomes
\begin{eqnarray}
\dot C_1 &=&  -i v C_2 - \kappa (E_1-E(t))^2\, C_1,
\nonumber\\
\dot C_2 &=&  -i v C_1 - \kappa (E_2-E(t))^2\, C_2.
\label{CnEqResonance}\end{eqnarray}
where $v=V_0/\hbar$.

If the measurement does not take place ($\kappa=0$), the equations describe Rabi oscillations, $C_1(t) =R_1(t)$, $C_2(t) =R_2(t)$, where
\ba
R_1(t)&=&C_1(0)\,\cos v t -iC_2(0)\,\sin v t, \\
R_2(t)&=&C_2(0)\,\cos v t -iC_1(0)\,\sin v t. 
\label{Rabi-osc}\ea
When the measurement does take place, the nature of the process will depend on the relative values of three parameters of the dimension of time: the duration of measurement $T$, the level resolution time $T_{\rm lr}$, and the Rabi period $T_R=\pi/v$. One can distinguish three characteristic regimes of measurement~\cite{AudMen97En}:

\quad

\noindent{\bf Zeno regime ($T_{\rm lr}\ll T_R\ll T$)}
\begin{itemize}
\item The measurement readout is $E(t)\simeq E_1$ or $E(t)\simeq E_2$.
\item The probability that $E(t)$ is close to $E_n$ is $|C_n(0)|^2$.
\item If $[E]$ is close to $E_n$, the system after the measurement is at level $n$.
\item Rabi oscillations are completely suppressed. 
\end{itemize}

\quad

\noindent{\bf Rabi regime ($T_R\ll T\ll T_{\rm lr}$)}
\begin{itemize}
\item Rabi oscillations remain unchanged. 
\item  $[E]$ is arbitrary in the band of the width ${\Delta E}_T={\Delta E}\sqrt{T_{\rm lr}/T}$. 
\end{itemize}

\quad

\noindent{\bf Intermediate regime ($T_R\sim T_{\rm lr} \sim T$)}
\begin{itemize}
\item The period of oscillations is slightly increased.
\item $[E]$ occurs in the band oscillating between the levels.
\item Oscillations of $[E]$ correspond to Rabi oscillations. 
\end{itemize}

The last regime is the most interesting, since it offers the possibility (with a certain degree of accuracy) of monitoring the quantum transition. We shall consider this regime in greater detail in Section~\ref{Sect-monitor}.

\subsection{Monitoring of quantum transition}
\label{Sect-monitor}

The set of equations \eq{CnEqResonance} was used in Ref.~\cite{AuMenNam97scat} for analyzing the continuous energy measurement in a two-level system occurring under the resonant force during the time interval $T_R/2=\pi/2v$ ($\pi$-pulse). Outside of this interval the potential \eq{resonance-potential}, as well as the coefficient $v$ in \Eq{CnEqResonance}, were assumed to be zero. In the absence of measurement, the resonant action would have led to the transition of the system from level 1 to level 2 with probability 1. One could expect that the measurement somewhat 
reduces the probability of transition, but allows monitoring continuously the state of the system in the course of transition.

The set of equations \eq{CnEqResonance} has been solved numerically for many functions $E(t)$ selected at random. For each function $E(t)$, calculated were the relevant functions $C_1(t)$ and $C_2(t)$, and the probability density $P[E]=|C_1(T)|^2+|C_2(T)|^2$ of the measurement readout $[E]$. The probability distribution of different measurement readouts was obtained, and the behavior of the system depending on the measurement readout analyzed. The behavior of the system is graphically represented by the function ${\cal P}_2(t)=|c_2(t)|^2$. In the absence of 
measurement this curve would go from zero to one over the duration of $\pi$-pulse. In the presence of measurement the curve may behave differently depending on the measurement readout $[E]$.

\begin{figure}
\let\picnaturalsize=N
\def\picsize{4.0in}
\def\picfilename{monitor.eps}
\ifx\nopictures Y\else{\ifx\epsfloaded Y\else\input epsf \fi
\let\epsfloaded=Y
\centerline{\ifx\picnaturalsize N\epsfxsize \picsize\fi \epsfbox{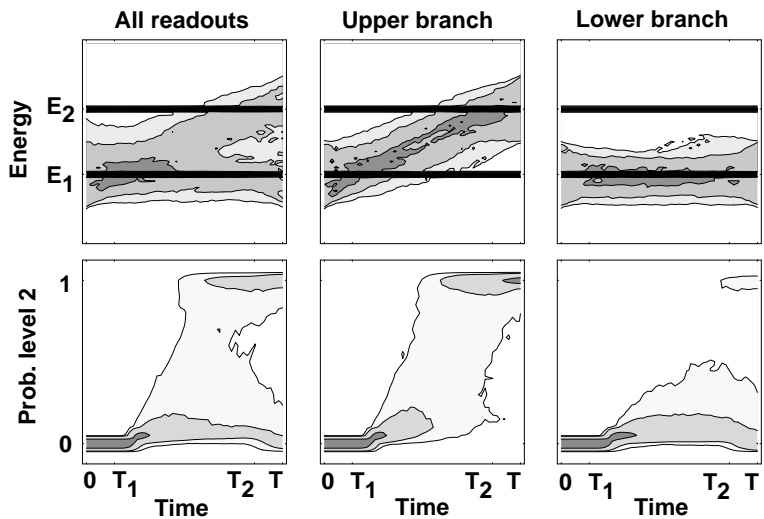}}}\fi
\caption{Monitoring of quantum transition with a moderately hard continuous measurement.}
\label{Monitor}\end{figure}

Figure~\ref{Monitor} shows the results of numerical calculation based on the set of equations \eq{CnEqResonance} for a moderately hard measurement. The density diagrams for curves $[E]$ (upper row) and $[{\cal P}_2]$ (lower row) indicate clearly which of these curves are more probable. It is very important to remember that curves $[E]$ have been smoothed before putting them on the diagram. Smoothing was performed in the time scale slightly smaller than the 
transition time $T_R/2$. It is because of such smoothing that the curves correctly represent the process of transition: curves $[E]$ without smoothing exhibit fast oscillations, and the information they carry is concealed.

The left-hand pair of diagrams in Fig.~\ref{Monitor} represents all possible measurement readouts. We see that they clearly split into two classes, in one of which the measurement readouts point to the presence of transition, while in the other the results indicate that the transition does not take place. The middle and right-hand pairs of diagrams represent these two classes of results separately. From these diagrams we see that there is correlation between the curves $[E]$ and $[{\cal P}_2]$ --- that is, the 
behavior of the system with a reasonable probability corresponds to the result obtained, although there is some probability of error (the measurement noise).

The general conclusions can be formulated as follows:
\begin{itemize}
\item The smoothed measurement readout $[E]$ with the probability of 80\% gives a correct description of the evolution of the system.
\item In the presence of measurement, the probability of transition decreases, and is close to 1/2 in the intermediate regime of measurement.
\end{itemize}

\subsection{Realization of continuous measurement of energy}
\label{SectRealization}

Practical implementation of continuous energy measurement in a two-level system was proposed in Ref.~\cite{AuMenNam97scat}. We take an isolated polarized atom, apply a $\pi$-pulse of resonant radiation to induce a transition between the levels, and measure the energy by shooting electrons one by one and looking whether the electron is deflected or continues in the initial direction. Scattering is caused by the dipole moment of the atom, which in turn depends on which level the atom occupies at the time. 
If the state of the atom is a superposition of states with a definite energy, the probability of scattering depends on the coefficients in the superposition, and thus on the mean energy of the atom in the given state. This in principle allows estimating the energy of the atom, whereas the back effect of the electrons on the atom results in the evolution of the atom described by the complex Hamiltonian (see Section~\ref{Sect-monitor}).

\begin{figure}
\let\picnaturalsize=N
\def\picsize{3in}
\def\picfilename{scatter.eps}
\ifx\nopictures Y\else{\ifx\epsfloaded Y\else\input epsf \fi
\let\epsfloaded=Y
\centerline{\ifx\picnaturalsize N\epsfxsize \picsize\fi \epsfbox{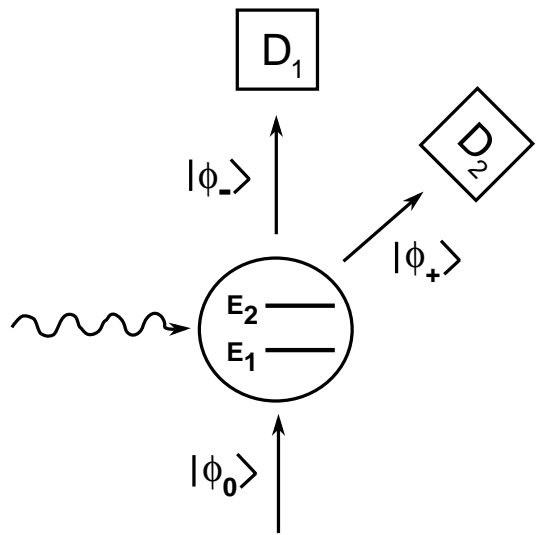}}}\fi
\caption{Short soft observation, repeated to implement the continuous energy measurement in a two-level system. $\Phi_0$ is the state of device before measurement, $\Phi_+$ ($\Phi_-$) is the state of device after positive (negative) result of observation. $D_+$ and $D_-$ are detectors. }
\label{scatter}
\end{figure}

This scheme of continuous measurement can be considerably generalized (this will be done in our subsequent paper \cite{AuMenEnGen}). In place of scattering of electrons by the atom, it is sufficient to arrange a long series of observations of a two-level system with an auxiliary measuring system (see Fig.~\ref{scatter}). The interaction involved in the observation must be weak, so as not to modify the state of the system to any considerable extent. Then the information gained through each 
observation in the series is also small, but a long series of observations will give information which is adequately represented with the aid of RPI or Schr\"odinger equation with complex Hamiltonian \eq{CnEqResonance}. The function $E(t)$, which represents the measurement readout in this equation, is constructed from experimental data in the following fashion.

Each observation may give either of the two possible results: positive, if the state of the measuring device after the interaction differs considerably from its initial state, or negative, if the end state of the measuring device is close to the initial state. A long series of observations is split into the shorter (but still long enough) chains of $N$ observations each. For each of these $N$-series we find the ratio $n=N_+/N$ of positive outcomes of scattering (number of deflected electrons) to the total number of 
electrons directed at the atom. This ratio is the experimental evaluation of the probability of scattering for the $N$-series in question. Since the probability depends on the mean energy, a simple calculation will also give an estimate for the mean energy $E$. As a result, we get one point on curve $[E]$. Performing the same procedure for each of the $N$-series, we obtain curve $E(t)$ which represents the 
result of continuous measurement and enters the equation with the complex Hamiltonian \eq{CnEqResonance}.

This train of arguments more or less supports the conclusion that a series of electron scatterings by the atom leads to a pattern predicted by the theory of continuous measurements --- that is, to the behavior represented by the solution of \Eq{CnEqResonance}. As a matter of fact, however, one can give a complete quantum mechanical treatment of this measuring configuration, and prove that there is full agreement between the predictions of the conventional quantum mechanical analysis and the 
results of the phenomenological RPI approach. This will be done in the forthcoming publication~\cite{AuMenEnGen}.

It is very interesting that a broad class of realizations results in the behavior of the system described by the same effective Hamiltonian with the additional quadratic imaginary term. The behavior of the measured system does not depend on the particulars of the measuring procedure --- it only depends on the single constant $\kappa$ (or $\Tlr$, which is equivalent) that appears as a certain combination of the parameters of the measuring system. This proves that the simple equation with 
the effective complex Hamiltonian, obtained through the RPI approach, is not a bad approximation. The behavior described by this equation is indeed characteristic of the real measuring systems.

\section{Conclusion}

In this paper we have considered the continuous measurement of a quantum system, which is a special dissipative process. The continuous measurement can be intended (as in case of repeated soft observations, Section~\ref{SectRealization}), or spontaneous (as in case of diffusion of particle, Section~\ref{QuDiffusion}). In any case, the continuous measurement is accompanied by gradual decoherence --- destruction of quantum superpositions of those states that are distinguished (resolved) by the measuring environment.

In addition, the continuous measurement may serve as a model of the process of decoherence associated with the measurement of a discrete variable (Section~\ref{MeasEnMultilevel}). If the description of such measurement disregards its finite duration, the process is adequately described by the von Neumann reduction postulate: instantaneous decoherence. When it is desirable to monitor the time evolution of decoherence, however, we come to deal with continuous measurement and gradual decoherence.

The continuous measurement of a quantum system gives specific features to its evolution, primarily the appearance of classical elements in addition to the quantum ones. Mathematically, such evolution can be described in different ways: using the Lindblad master equation, the stochastic Schr\"odinger equation, or the equation with the complex Hamiltonian for the state vector (Section~\ref{Sect-phenomen}). In the last case, the 
imaginary term of the Hamiltonian, which accounts for the effects of the measuring medium, depends on the information obtained in the course of continuous measurement. This is of utmost importance, being a demonstration of the dynamic role of information~\cite{Men97rev}.

The continuous measurement can be softer or harder, depending on the strength of interaction with the measuring medium (mathematically, this is determined by the magnitude of the imaginary term in the Hamiltonian). Soft measurement has little effect on the dynamics of the measuring system, but the information about the system is small. Hard measurement gives more information but considerably modifies the dynamics. In the limit of very hard measurement of the discrete 
observable we deal with Zeno effect: the system is frozen, the transitions between the states with different values of the observable become unlikely (Section~\ref{MeasEnMultilevel}).

The most interesting is the intermediate regime of measurement, when the measurement is not hard enough for freezing the dynamics of the system completely, but is sufficiently hard for gaining a reasonable amount of information. In this regime it becomes possible to monitor the quantum transition, although such information is gained at the expense of reducing the probability of transition (Section~\ref{Sect-monitor}). The continuous measurement permitting the monitoring of a quantum 
transition can be realized as a series of soft short interactions of the two-level system with some auxiliary system (Section~\ref{SectRealization}). One possible realization employs a series of electron scatterings by the atom.

Disputes on conceptual matters have never ceased since the early days of quantum mechanics. The root of these problems is the incompatibility of the classical and quantum descriptions of physical systems and processes. This incompatibility is most dramatically manifested in the measurements aimed at obtaining classical information about a quantum system. Today we are able to study in detail the behavior of the measured system (and the measuring system if 
necessary) in the course of measurement, and this is the subject of this paper (see also the excellent review \cite{Zeh-bk96}). The conceptual problems of the theory of measurement, however, remain essentially unresolved.

Simplifying the situation to the utmost, we may say the following. The measurement is associated with the selection of one out of many alternatives. We know well what happens when a particular alternative is selected, and can calculate the probability of each alternative. This gives answers to all questions that may arise in practice. How and why the selection is made, however, remains obscure. This question is hard to answer because this requires reconciling the quantum and the 
classical visions of the world. The attempts at solving this problem lead to very unusual constructions, the many-worlds interpretation of quantum mechanics being one example \cite{Everett73}. This scope of problems that may be regarded as the conceptual aspect of decoherence has been left out completely in the present paper.

\quad

\centerline{\bf Acknowledgement}

The author is thankful to many colleagues for fruitful discussions clarifying  questions of quantum mechanics. Especially acknowledged are discussions with J.~Audretsch, V.~A.~Namiot and H.-D.~Zeh. This work was supported in part by the Russian Foundation of Basic Research, grant 98-01-00161



\end{document}